\documentclass[12pt,preprint]{aastex}
\usepackage{psfig}

\slugcomment{To appear in the Astrophysical Journal}
\shorttitle{Chandra Spectroscopy of SS 433}
\shortauthors{Marshall et al.}

\begin{document}

\title{The High Resolution X-ray Spectrum of SS 433 using
the {\em Chandra} HETGS}

\author{Herman L. Marshall, Claude R. Canizares, and
    Norbert S. Schulz}
\affil{Center for Space Research, MIT, Cambridge, MA 02139}
\email{hermanm@space.mit.edu,
crc@space.mit.edu,
nss@space.mit.edu}

\begin{abstract}

We present
observations of SS 433 using the Chandra High Energy Transmission
Grating Spectrometer.  Many emission lines of highly ionized
elements are detected with the relativistic blue and red Doppler shifts.
The lines are measurably broadened to 1700 km s$^{-1}$
(FWHM) and the widths do not depend significantly on the characteristic
emission temperature, suggesting that the emission occurs in a freely
expanding region of constant collimation with opening angle of
1.23 $\pm$ 0.06\arcdeg.
The blue shifts of lines from low temperature gas are the same as those of
high temperature gas within our uncertainties, again
indicating that the hottest gas we observe
to emit emission lines is already at terminal velocity.
This velocity is 0.2699 $\pm 0.0007 c$, which is larger than the
velocity inferred from optical emission lines by 2920 $\pm$ 440
km s$^{-1}$.
Fits to the emission line fluxes give a range of temperatures
in the jet from $5 \times 10^6$ to $1 \times 10^8$ K.
We derive the emission measure as a function of temperature
for a four component model that fits the line flux data.
Using the density sensitive Si {\sc XIII} triplet, the
characteristic electron density is $10^{14}$ cm$^{-3}$ where the
gas temperature is about $1.3 \times 10^7$ K.
Based on an adiabatic expansion model of the jet and a
distance of 4.85 kpc, the electron densities drop
from $\sim 2 \times 10^{15}$ to $4 \times 10^{13}$ cm$^{-3}$
at distances of $2-20 \times 10^{10}$ cm from the apex of the
cone that bounds the flow.  The radius of the
base of the visible jet is estimated to be $\sim 10^8$ cm
and the mass outflow rate is $1.5 \times 10^{-7} ~ M_{\sun}$
yr$^{-1}$.  The kinetic power is $3.2 \times 10^{38}$ erg s$^{-1}$,
which is $\times$1000 larger than the
unabsorbed 2-10 keV X-ray luminosity.
The bremsstrahlung emission associated with the lines can
account for the entire continuum; we see no direct evidence
for an accretion disk.
The image from zeroth order shows
extended emission at a scale of $\sim$2\arcsec, aligned in the
general direction of the radio jets.

\end{abstract}


\section{Introduction}

The Galactic X-ray binary SS 433 is the only known astrophysical system
where relativisticly
red- and blue-shifted lines are observed from high Z elements.
Lines from the H Balmer series were detected by \citet{margon77} and
modelled kinematically by \citet{am79}, \citet{fr79},
and \citet{milgrom79} as emission from opposing
jets emerging from the vicinity of a compact object that
precesses about an axis inclined to the line of sight.  The precise
description of the line positions was given by \citet{ma89}: the jet
velocity, v$_j$, is 0.260$c$ and its orientation precesses with a
162.5 day period in a cone with half-angle 19.85\arcdeg\ about
about an axis which is 78.83\arcdeg\ to the line of sight.
The lines from the jet are Doppler shifted
with this period so that the maximum
redshift is about 0.15 and the maximum blueshift is about -0.08.
The X-ray source undergoes eclipses at a 13.0820 day period.
Assuming uniform outflow,
the widths of the optical lines were used to estimate the
opening angle of the jet: 5\arcdeg\ \citep{begelman}.
For more details of the optical spectroscopy, see the review by
\citet{margon84}.
 

SS 433 has radio jets oriented at a position angle of
100\arcdeg\ (east of north) which show an oscillatory pattern
that can be explained by helical motion of material flowing
along ballistic trajectories \citep{hjellming}.
Using VLBI observations of ejected knots, \citet{vermeulen}
independently confirmed the velocity of the jet and
the kinematic model and also derived
an accurate distance: 4.85 $\pm$ 0.2 kpc, which we adopt
throughout our analysis.
The radio jets extend from the milliarcsec scale to several
arcseconds, a physical range of $10^{15-17}$ cm from the
core.  The optical emission lines, however, originate
in a smaller region, $< 3 \times 10^{15}$ cm across, based
on light travel time arguments \citep{dm80}.
At the same position angle is large scale X-ray emission that
extends $\pm$ 30-35\arcmin\ to about $1.5 \times 10^{20}$
cm \citep{watson83}.

\cite{marshall79} were the first to demonstrate that SS 433 is an X-ray
source.  Their {\em HEAO-A} spectrum could be
modelled as thermal bremsstrahlung
with $kT = 14.3$ keV and emission due to Fe-K was detected near 7 keV.
They also found that the X-ray luminosity, $\sim 10^{36}$ erg s$^{-1}$,
was much lower than the inferred power in kinetic energy:
$\sim 10^{40}$ erg s$^{-1}$.
{\em EXOSAT} X-ray observations showing that the 6.7 keV line
appeared to shift with the precession phase were interpreted as
detecting Fe {\sc XXV} from 10 keV gas \citep{watson}.
Using {\em Ginga}, \citet{brinkmann91}
concluded that $kT$ should be 30 keV while
\citet{kotani} obtained a value of 20 keV based on the ratio of 
Fe {\sc XXV} and Fe {\sc XXVI} line fluxes from {\em ASCA}
observations.  \citet{brinkmann91}
and \citet{kotani} developed the model that
the X-ray emission originates near the base of the jets which are
adiabatically expanding and cooling until $kT$ drops to
about 100 eV at which point the jet is thermally unstable.
\cite{kotani} found that the redshifted Fe {\sc XXV} line
was fainter relative to the blueshifted line, than expected
from Doppler intensity conservation, leading them to
conclude that the redward jet must be obscured by neutral material
in an accretion disk.  Although the {\em ASCA} spectra showed
emission lines that were previously unobserved, the lines were
not resolved spectrally and lines below 2 keV were difficult
to identify.

SS 433 was observed using the High Energy Transmission
Grating Spectrometer (HETGS)
in order to resolve the X-ray lines, detect fainter lines than
previously observed and measure lower energy lines that could not
be readily detected in the {\em ASCA} observations.  A description of
the HETGS and its performance are given by \cite{canizares00}.
An important feature of the HETGS is that spectra are obtained
independently and simultaneously using the High Energy Gratings
(HEGs) and the Medium Energy Gratings (MEGs).  The HEGs have
$\times 2 $ higher spectral resolution and greater effective
area above 4 keV than the MEGs
while the MEGs have an extended bandpass and more effective area
for detecting low energy emission lines.

\section{Observations and Data Reduction}

SS 433 was observed with the HETGS on 23 September 2000
(JD - 2451000. = 445.0 - 445.4).  The
precession phase was 0.51, based on the ephemeris determined
by \citet{ma89} and the orbital phase was 0.67 using the orbital
ephemeris of \cite{gladyshev}.
The telescope roll angle was set so that the dispersion
direction was nearly perpendicular to the jet axis.
The exposure time was
28676 s, determined by $t_{exp} = N (t_{frame} - t_{shift})$,
where $N$ is the number
of frames with detected events, $t_{frame}$ is the ACIS-S frame
time (3.241 s) and $t_{shift}$ is the frame shift time (0.041 s).
This computation corrects for the ``readout streak'' exposure because
the streak data are not included in the analysis.

\subsection{Imaging}

Based on the HETGS spectral fits (see sec.~\ref{sec:spectra}), we
expected to observe about 1.4 count/s in zeroth order.
The average count rate in zeroth order was measured to be about 0.15 count/s,
so we conclude that the image is heavily affected by pileup,
given the 3.2 s frame times.
Briefly, pileup occurs when two or more events are detected within
an event region (3 $\times$ 3 pixels) in a readout or frame time.
Pileup has three important effects: 1) event energies are combined
so that it is impossible to determine the energies of the original
events, 2) the event grade (denoting the distribution of charge
among the 9 pixels) changes and can often end up as a
grade that is normally eliminated in processing, reducing the
apparent count rate, and 3) having a highly nonlinear dependence on
the local count rate, the cores of point sources are more affected
than the wings so the appearance can be distorted.  Spectral analysis
of heavily piled zeroth order data can be challenging and was deemed to be of
limited value compared to the HETGS spectra.

\begin{figure}
\plotone{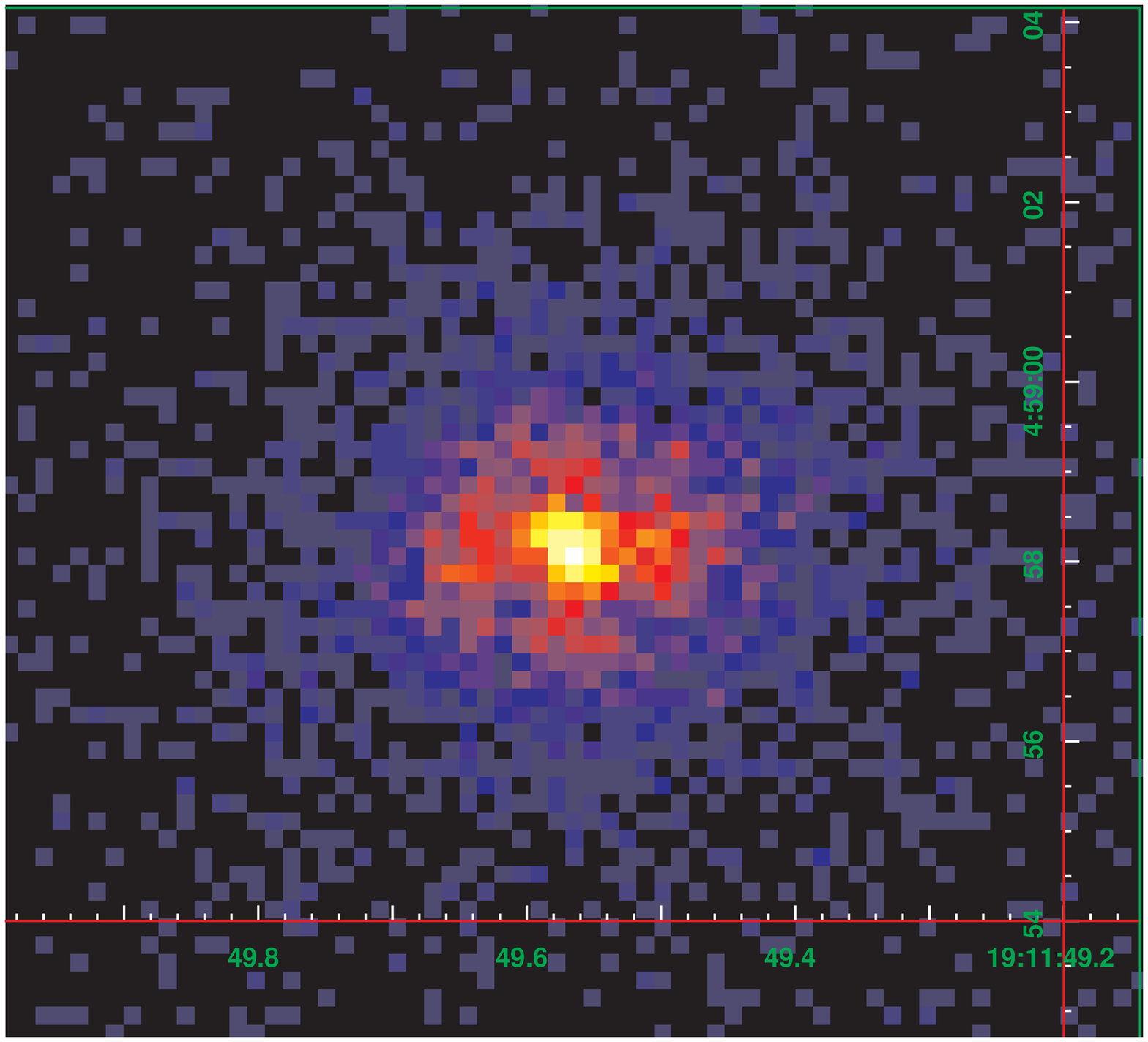}
\caption{Image of SS 433 from the zeroth order of the
{\em Chandra} High Energy Transmission Grating Spectrometer.
The dispersion direction is approximately aligned
with the north-south direction.
The image is extended along the east-west direction on a scale
of 2-5\arcsec.  The extent is comparable to that observed in
the radio band \citep{hjellming}.
\label{fig:image} }
\end{figure}

\begin{figure}
\plotone{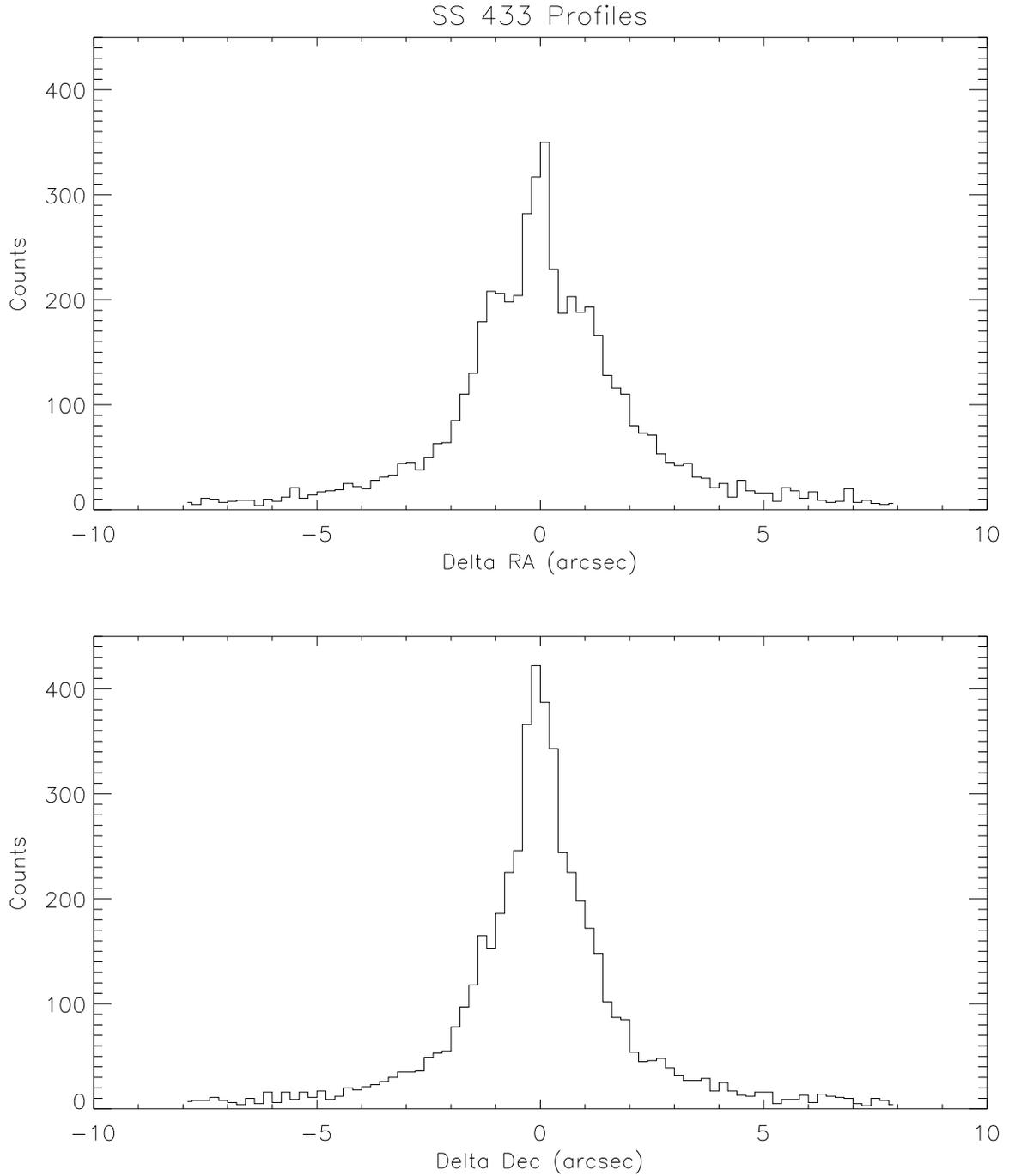}
\caption{One dimensional profiles of the zeroth order image
shown in figure~\ref{fig:image}.  {\em Top:} East-west
direction, which is nearly aligned with the radio jets
observed by \citep{hjellming}.  {\em Bottom:} North-south
direction.  The difference between the N-S and E-W profiles
is significant and is independent of the distortions of the
point response that occurs due to event pileup (see text).
\label{fig:1dprofiles} }
\end{figure}

The zeroth order image, shown in Fig.~\ref{fig:image}, is distinctly
elongated in the E-W direction.  Due to event pileup, we cannot determine
if the image matches that of a point source within 1\arcsec\ of the core.
Profiles along the E-W and N-S directions should be affected by
pileup in the same way, however we find that they are clearly different,
as shown in Fig.~\ref{fig:1dprofiles}: there are ``shoulders'' to the
E-W profile that do not appear in the N-S profile.  Again, these
profiles are difficult to compare to point source models due to the
nonlinear effects of pileup, so these one-dimensional profiles are most
useful only for comparing to each other.

\begin{figure}
\plotone{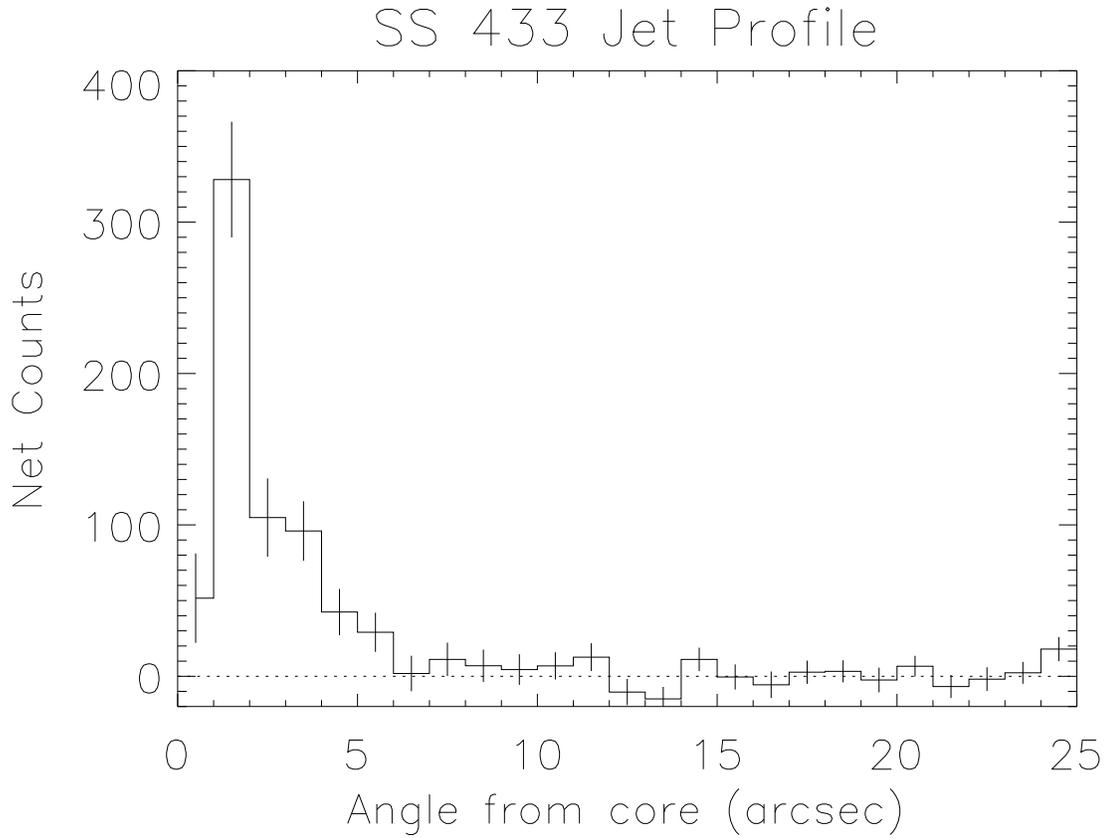}
\caption{Profile of the 2\arcsec\ scale X-ray jet.
North and south sectors were used as background so that the
core point source is eliminated from the east-west sectors
and the flux in the first bin is nulled.
The jets are brightest at about 1.5\arcsec\ from
the core and are no longer detectable beyond 6\arcsec.
\label{fig:radialprofile} }
\end{figure}

We associate the excess emission in the E-W direction
with the arcsec-scale radio jet, as observed by \citet{hjellming}
using the VLA.  A radial profile of the extended
emission was computed by determining the counts in two E-W
azimuthal sectors and subtracting normalized profiles using the remaining
N-S azimuths.  The profile is shown in Fig.~\ref{fig:radialprofile}.
This method nulls the point source, so the 1\arcsec\ bin centered on
the source shows no significant excess flux.  Extended flux is detected
1-2\arcsec\ from the core and
drops rapidly with distance from the image centroid until it is
undetectable beyond 6\arcsec.  We estimate that the total count rate
from the extended jets is about 0.008 $\pm$ 0.0005 count/s, accounting
for about 0.6\%
of the flux within 5\arcsec\ of the source centroid.

A set of cross dispersion profiles were derived for continuum
regions and for several line regions of the dispersed spectrum.
For comparison, profiles for
the same wavelength regions were determined from a point-like calibration
source, PKS~2155--304 (observation ID 1705).
The profiles are indistinguishable from
each other and are consistent with that of the point source.
Thus, we conclude that the most of the emission lines and
the continuum originate from the same region on a scale
of $<$ 1\arcsec.  The extended, elongated emission detected in the
zeroth order image would not be
discernable in the cross dispersion profiles at the same low
fractional power due to the slight astigmatism of the
grating optics in the cross dispersion direction.

\subsection{Spectra}
\label{sec:spectra}

The spectral data were reduced starting from level 1 data provided
by the Chandra X-ray Center (CXC) using
IDL custom processing scripts; the method is quite similar to
standard processing using standard CXC software but was developed
independently for HETGS calibration work.
The procedure was to: 1) restrict
the event list to the nominal grade set (0,2,3,4, and 6), 2) remove the
event streaks in the S4 chip (which are not related to the readout streak),
3) estimate exposure including accounting for frame shift
time, 4) determine the location of zeroth
order using Gaussian fits to one dimensional profiles, 5) rotate
events from sky coordinates to compensate for the telescope roll, 6) compute
the dispersed grating coordinates ($m\lambda$ and $\phi$)
using the grating dispersion angles
and the dispersion relation, 7) correct event energies for detector
node-to-node gain variations, 8) select $\pm 1$ order events
using $| E_{ACIS} * m \lambda / (h c) | - 1 < \Delta $ where $E_{ACIS}$
is the event energy inferred from the ACIS pulse height, and $\Delta$
is 0.20 for the MEG and 0.15 for the HEG events, 9) eliminate events
in bad columns where the counts in an histogram deviated by 5$\sigma$
from a 50 pixel running median, 10) select ``source'' events
spatially within a rectangular region
3.6\arcsec\ of the center of the dispersion line and
background events in the bands 7.2 to 21.6\arcsec\ from the dispersion
line, 11) bin MEG (HEG) events at 0.01\AA\ (0.005\AA), 12) eliminate
data affected by detector gaps, and 13) generate
and apply an instrument effective area based on the pre-flight calibration
data.

\begin{figure}
\epsscale{0.8}
\plotone{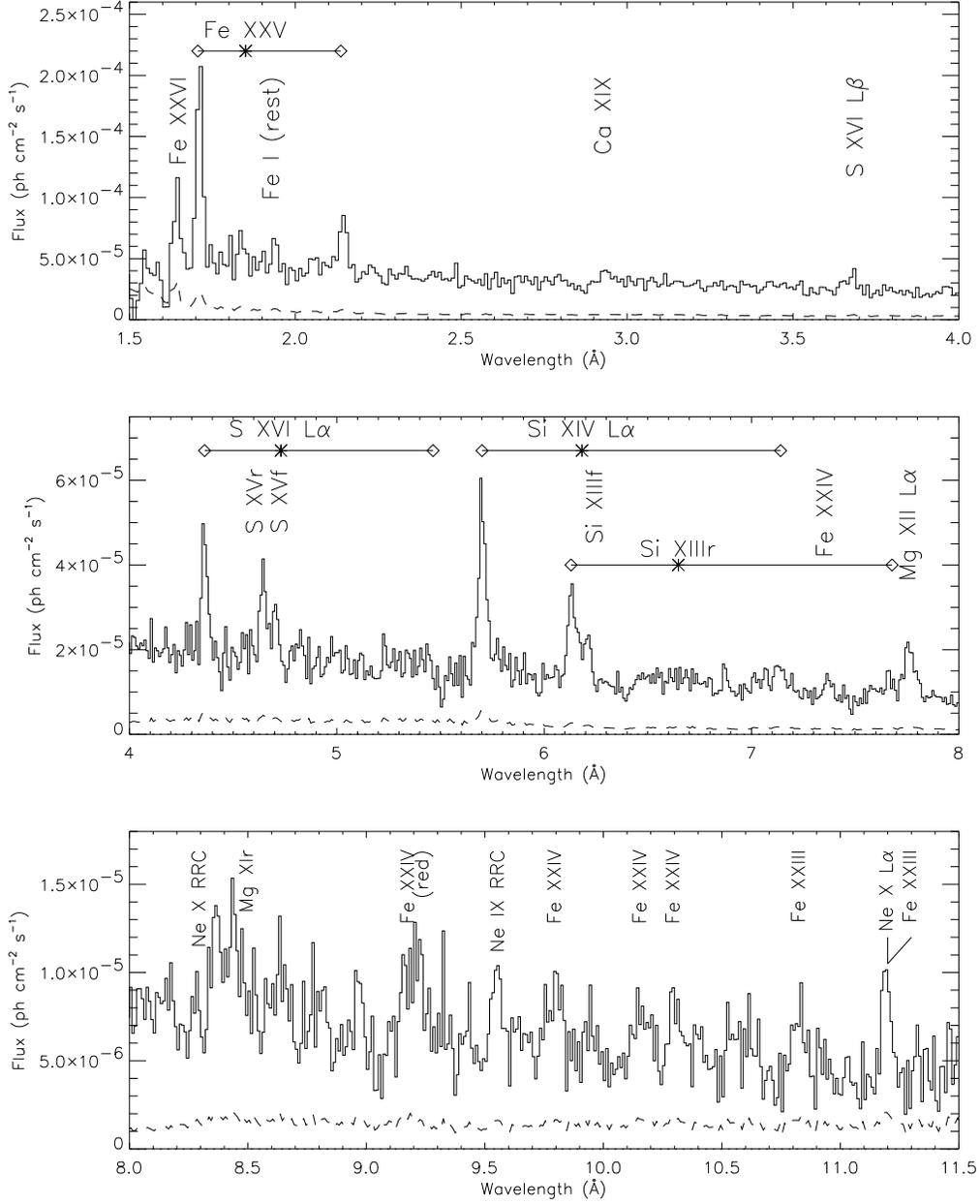}
\caption{The X-ray spectrum of SS 433 observed with the
Chandra High Energy Transmission Grating Spectrometer.
The HEG and MEG data were combined at the resolution
of the MEG spectrum.
Line identifications are shown and measurements are
given in table~\ref{tab:linefluxes}.
All lines originate in the blue jet unless shown otherwise.
Horizontal lines connect the locations of blue- and red-shifted
lines (diamonds) to the rest wavelengths (asterisks).
The dashed line gives the statistical uncertainties.
Emission lines are resolved and their strengths indicate that the
plasma is collisionally dominated.
The Si {\sc XIII} triplet is resolved in the HEG data,
as shown in Fig.~\ref{fig:si13}, and can be used to estimate
the density of the emission region.
\label{fig:spectrum} }
\end{figure}

\begin{figure}
\epsscale{0.8}
\plotone{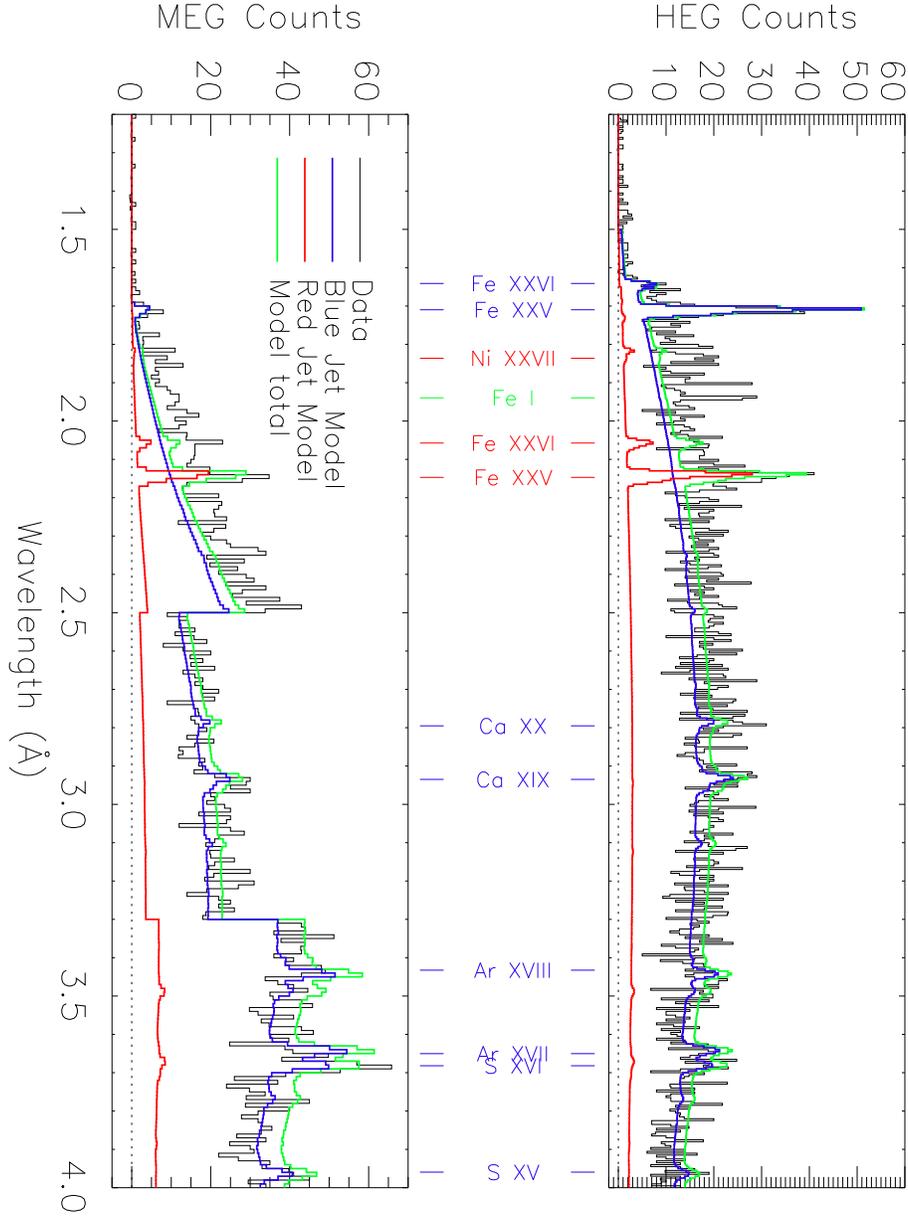}
\caption{The 1.2-4.0 \AA\ portions of the HEG and MEG
spectra of SS 433 observed with the
Chandra HETGS, compared to models of the spectra of the blue
and red jets.  The sum of the red and blue spectra give the
green curve.
Line identifications are shown and measurements are
given in table~\ref{tab:linefluxes}.
The continuum is dominated by thermal bremsstrahlung emission.
Edges in the MEG spectrum at 2.5\AA\ and 3.3\AA\ are the result
of excising data near the detector chip gaps (see text).
\label{fig:spectra1} }
\end{figure}

\begin{figure}
\epsscale{0.8}
\plotone{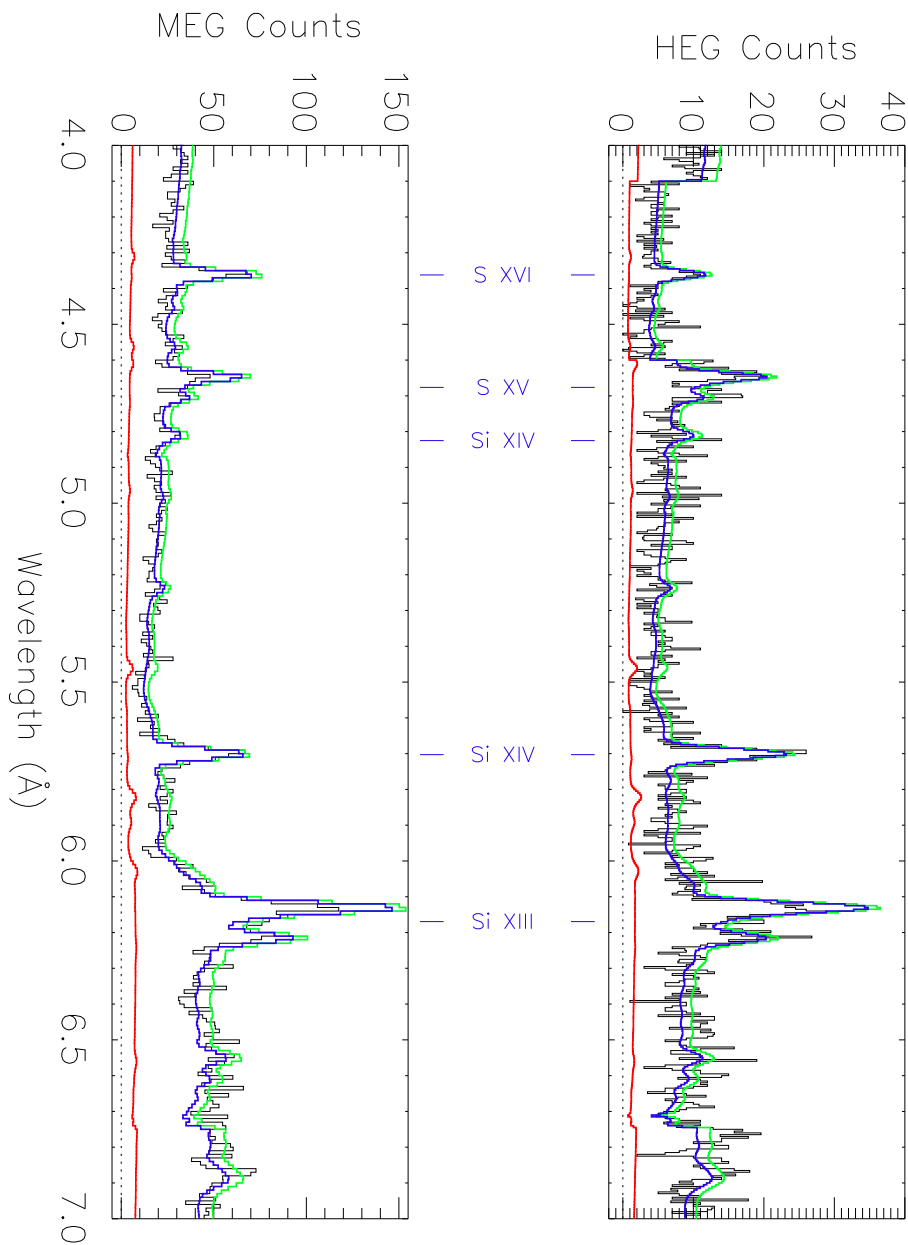}
\caption{Same as figure~\ref{fig:spectra1} except for the 4-7\AA\ region.
Lines from the red jet are not very strong in this portion of the
spectrum.  The overall model continuum is slightly higher than the
data in the 4-6\AA\ range.  Due to limitations of the
modelling code, density-sensitive emission lines are
not accurately represented.  A better model of the Si {\sc XIV}
and Si {\sc XIII} lines is shown in detail in Fig.~\ref{fig:si13}. 
\label{fig:spectra2} }
\end{figure}

\begin{figure}
\epsscale{0.8}
\plotone{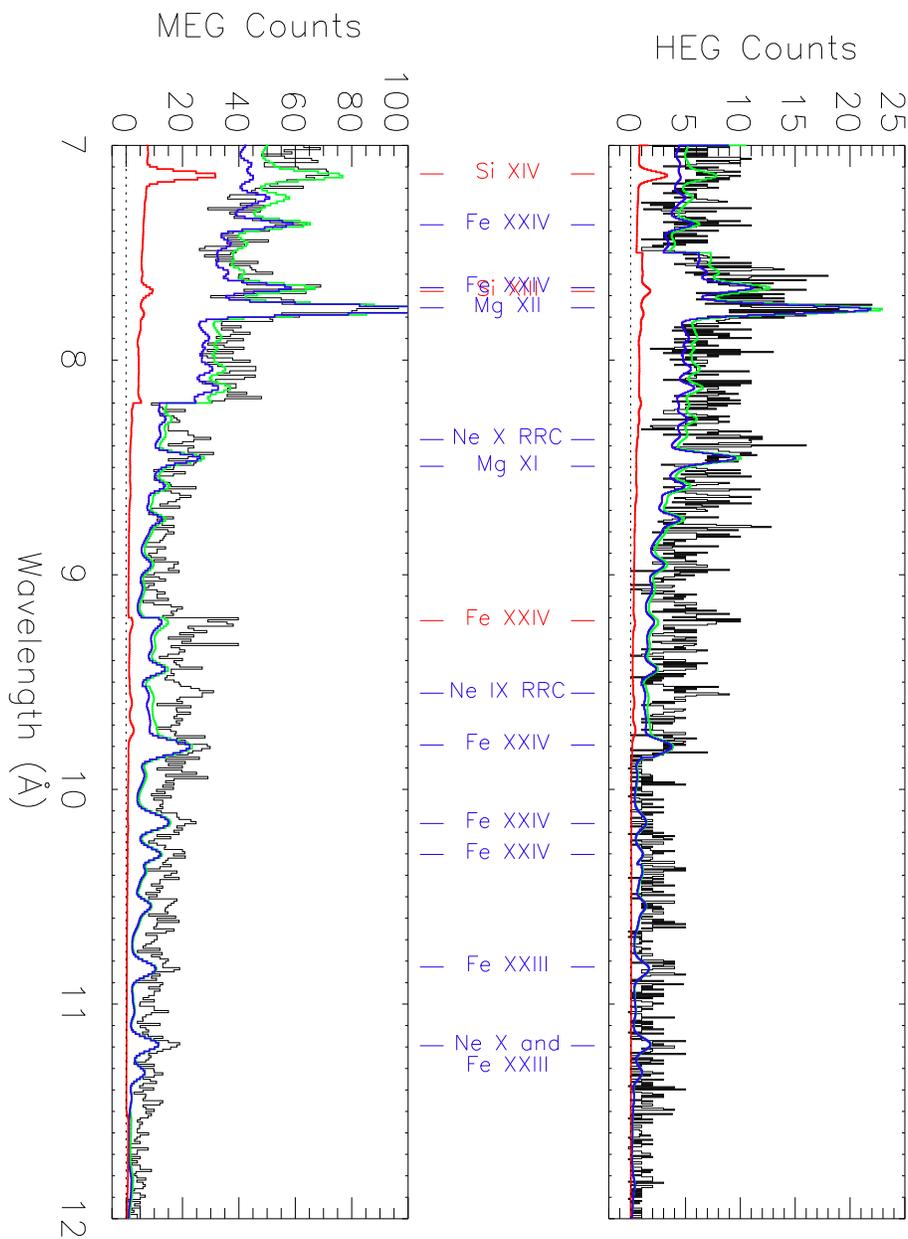}
\caption{Same as figure~\ref{fig:spectra1} except for the 7-12\AA\ region.
The overall model continuum is slightly lower than the
data in this wavelength range.  Two blueshifted radiative recombination
continuum (RRC) features are marked.  The RRC features are
not predicted by the model which assumes purely collisional
ionization balance.
\label{fig:spectra3} }
\end{figure}

The spectra are shown in Figs.~\ref{fig:spectrum} to \ref{fig:spectra3}.
Figure~\ref{fig:spectrum} shows the flux corrected spectrum, combining
the MEG and HEG data with statistical weighting.
There are many broad emission lines and a significant continuum.
At the low energy end, the lines are so broad that it becomes
difficult to discern them from the continuum.
A simple power law fit to the continuum gave a very good fit
to the 0.8-8 keV spectral data:
$n(E) = A E^{-\Gamma} e^{-N_H \sigma(E)}$, where n(E) is the photon
flux in ph cm$^{-2}$ s$^{-1}$ keV$^{-1}$, $E$ is energy in keV, $N_H$
is the line of sight neutral hydrogen density due to the
interstellar medium (ISM) primarily, $\sigma(E)$ is the
opacity of the ISM with cosmic abundances, and the best fit parameters
are $A$ = 0.015 ph cm$^{-2}$ s$^{-1}$ keV$^{-1}$, $\Gamma$ = 1.35,
and $N_H = 9.5 \times 10^{21}$ cm$^{-2}$.  The HEG and MEG spectra
give comparable results.  The $N_H$ and $\Gamma$ are about 30\% higher
but $A$ is 30\% lower than that found by \citet{kotani} although the
observations were obtained at the same precession phase.
The average 2-10 keV luminosity is about $3.1 \times 10^{35}$
erg s$^{-1}$ and was observed to decrease about 10\% during the
observation.

Line fluxes are given in table~\ref{tab:linefluxes}.
Lines were fit to the MEG and HEG data jointly
in six wavelength ranges after subtracting
a polynomial fit to the
continuum.  To allow for slight deviations from the continuum
model, a constant level was allowed to vary in each interval.
The line positions and fluxes were allowed to vary from initial
estimates for each line but the Gaussian widths were fitted to
a single value in each range.  The widths are given in
table~\ref{tab:linewidths}.  Uncertainties were estimated
upon fixing the parameters of all other lines in a given interval
to their best fit values.  Only statistical errors are quoted;
systematic uncertainties other than possible line misidentifications
are expected to be smaller than 10\%.

\begin{figure}
\epsscale{0.8}
\plotone{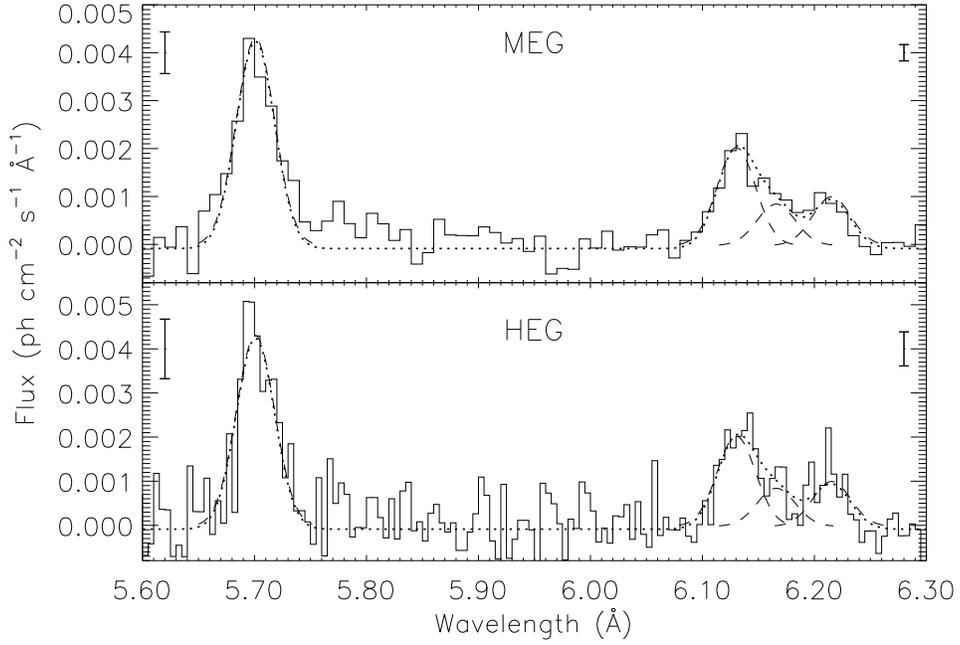}
\caption{Detail of Gaussian modelling of the Si {\sc XIV}
and Si {\sc XIII} lines.
The dotted line gives the combined spectral model and
the components are shown with dashed lines.  Two typical
error bars are shown in each panel.
The Gaussian dispersions were constrained to be identical for
all lines and the rest wavelengths of all lines were fixed
so there were only six parameters in the model: Doppler shift,
line width, two line strengths (Si {\sc XIV} Ly$\alpha$ and
Si {\sc XIII} resonance), and the ratios $G = (i+f)/r$
and $R = f/i$, where $r$, $i$, and $f$ are the strengths
of the Si {\sc XIII} resonance, intercombination, and forbidden lines, 
respectively.  A simple continuum model was subtracted from
the spectrum.
\label{fig:si13} }
\end{figure}

A redshift was measured for each line or line triplet after
determining the ID based on the expected shifts of the blue
and red jets from the kinematic model.  For the precession
phase of our observation, the predicted blue- and red-shifts
were -0.0670 and 0.1384, using the equation and parameters in
table 1 of \cite{ma89}.
Wavelengths of line blends were obtained using wavelengths
from APED\footnote{See {\tt http://hea-www.harvard.edu/APEC/aped/general.html}
for more information about APED.} weighted by
the relative fluxes of each component, which was especially
important for the He-like triplets.  The S {\sc XV} and Si
{\sc XIII} triplets were treated somewhat differently.
We determined more precise redshifts for these
blends fitting the line profiles with several Gaussian
components of variable widths but we fixed the rest wavelengths to
the values given by APED, as implemented in {\tt ISIS}\footnote{See {\tt http://space.mit.edu/CXC/ISIS} for more information about ISIS.}.
Fig.~\ref{fig:si13} shows the results of jointly fitting
the MEG and HEG data.  The Si {\sc XIII} line is used to estimate
the density in the jet in section~\ref{sec:models}.
In the red jet system, it was too difficult to measure
an accurate redshift for the Si {\sc XIII} triplet
due to blending
with Mg {\sc XII} from the blue jet and to its low flux.
The identified lines
were then grouped according to redshift.  Only one line
was not assigned to either the red or blue jets: a neutral
Fe-K line at rest in the observed frame.  It is unresolved
at HEG resolution, indicating a FWHM $<$ 1000 km s$^{-1}$.

\section{Line Widths and Positions}

Table~\ref{tab:linefluxes} shows that
the Doppler shifts of the lines in the blue jet system are
consistent with a single velocity to within the uncertainties,
as are those in the red jet system, although there are very
few positively identified lines in the red jet.
There are a few deviations
that are not likely to be indicative of intrinsic variations.
The lines generally responsible for the largest
deviations are mostly weak, such as Ca {\sc XX}, or
are weak He-like triplets, such as Ar {\sc XVII} and Mg {\sc XI}.
Furthermore, all deviations are substantially smaller than
the observed line widths, which are of order 0.006 (FWHM).
We determine that the Doppler
shift of the blue jet was $z_b = -0.0779 \pm 0.0001$ during this
observation.  For the red jet, the unblended
lines detected at 3$\sigma$ or better give an estimated
redshift of $z_r = 0.1550$ with a formal uncertainty of 0.0004.
There is marginal evidence that the Fe {\sc XXV} line
has a higher redshift than the other three lines.  The difference
is small, $\sim 800$ km s$^{-1}$, and could instead result from
underestimated uncertainties because the two faintest
lines are only
marginally detected and the Fe {XXIV} line could even be misidentified.
No other lines from the red jet
show positive velocity deviations except Si {\sc XIII}, which
is severely blended with the Mg {\sc XII} line from the blue jet.
The spectral shifts are comparable to values obtained from a
model fitted to measurements of the H-$\alpha$ lines
taken from the same precession cycle (Douglas Gies, private
communication) within the large intrinsic variations.

Assuming perfectly opposed jets at an angle $\alpha$ to the
line of sight, the Doppler shifts of the blue and red jets
are given by

\begin{equation}
\label{eq:redshift}
z = \gamma (1 \pm \beta \mu ) - 1
\end{equation}

\noindent
where v$_j = \beta c$ is the velocity of the jet flow,
$\gamma = (1 - \beta^2)^{-1/2}$, and $\mu = \cos \alpha$.
The blue (red) Doppler shift is obtained by using the $-$ ($+$) sign.
With high accuracy redshifts, we can obtain an estimate of
the $\gamma$ and $\beta$ by adding the redshifts to cancel the
$\beta \mu$ terms:

\begin{equation}
\label{eq:gamma}
\gamma = \frac{z_b + z_r}{2} + 1
\end{equation}

\noindent
giving $\beta = 0.2699 \pm 0.0007$.  Comparing to the
value derived by \cite{ma89} for the H-$\alpha$ lines,
the velocities of the
X-ray lines are larger by 2920 $\pm$ 440 km s$^{-1}$.
In order to reduce our estimate of $\beta$
to within 3$\sigma$ of the value derived from the H-$\alpha$
lines, the more uncertain X-ray Doppler shift -- the redshift --
would have to be as small as 0.1520.
A redshift this small would place the centroid
of the best measured line, Fe {\sc XXV}, practically outside
the observed line, so we have some confidence in the
jet velocity measurement.
Substituting our value for $\beta$ back into
Eq.~\ref{eq:redshift} and solving for $\alpha$ gives the
angle of the jet to the line of sight during our observation:
$\alpha = $ 65.46\arcdeg $\pm$ 0.07\arcdeg.

All jet lines are clearly resolved.
The line widths in table~\ref{tab:linewidths} are
consistent with the weighted average value ($\sigma$)
of 729 $\pm$ 34 km s$^{-1}$ or FWHM = 1710 $\pm$ 80 km s$^{-1}$.
There is only marginal evidence for a trend that the lower energy
lines are slightly narrower than average.
The widths of the red jet lines are consistent with that of the blue
jet.

\begin{figure}
\epsscale{1.0}
\plotone{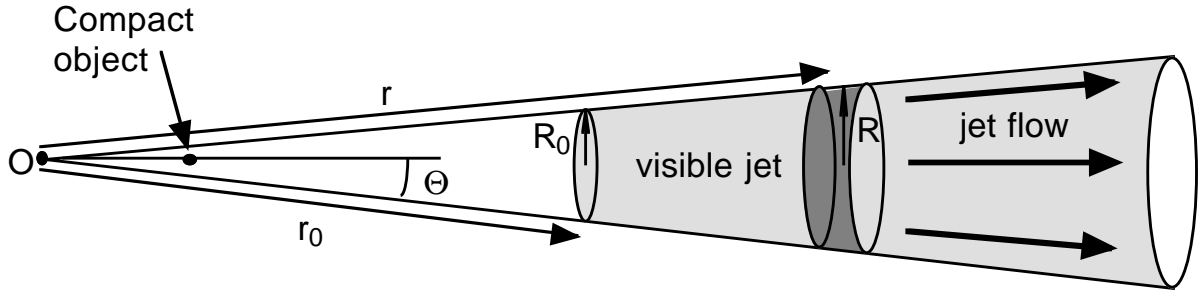}
\caption{Geometry of a jet with uniform velocity outflow.
The jet flow is radial, so quantities such as density and temperature
can be given in terms of $r$, the distance from the cone's apex at $O$.
The visible portion of the jet is shaded and its base is at a distance $r_0$
from the apex.  The compact object is on the jet axis at some
distance less than $r_0$ from $O$.
The radius of the jet cross section is $R = \Theta r$
and the radius at the base is $R_0 = \Theta r_0$.
\label{fig:jetgeometry} }
\end{figure}

The line widths are too large to result from
thermal broadening for $kT < 10$ keV (see sec.~\ref{sec:models})
-- 100-200 km s$^{-1}$ --
so we ascribe the widths to Doppler broadening that would result
from the divergence of a conical outflow \citep{begelman}.
Fig.~\ref{fig:jetgeometry}
shows the jet geometry, defining $r$ as the distance along
the cone from the apex and $R = \Theta r$ is the jet cross
sectional radius at $r$.
We assume that the density is uniform through the
cone's cross section and that the component of the velocity
that is parallel to the jet axis is
the same for all fluid elements in the slice.
Differentiating Eq.~\ref{eq:redshift} with respect to $\alpha$ gives
the line width at zero intensity for material flowing within a small angle
$\Delta \alpha$ of the jet axis: $\Delta z =
\gamma \beta \sin \alpha \Delta \alpha$.
Assigning $\Delta \alpha = \Theta$ gives $\Theta$ in terms
of v$_m = c \Delta z$, the maximum velocity relative to line center:

\begin{equation}
\Theta = \frac{ {\rm v}_m}{\gamma \beta c \sin \alpha}  .
\end{equation}

\begin{figure}
\epsscale{0.5}
\plotone{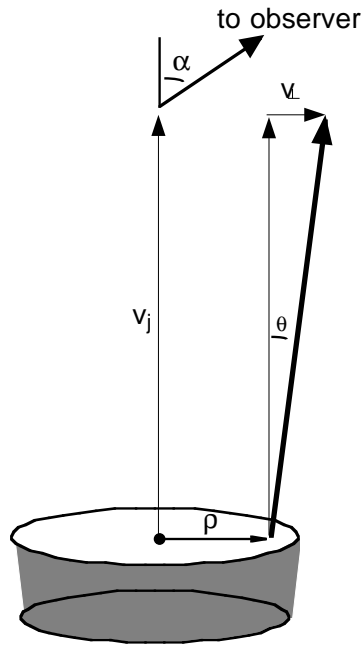}
\caption{Geometry of a differential slice of the jet, indicating
the various angles in the system.  The jet axis is viewed at an
angle $\alpha$ to the line of sight and $\theta$ is the flow
direction of a jet fluid element relative to the jet axis.
Gas with constant line-of-sight
velocity, v, relative to the on-axis flow, lies along a chord at distance
$\rho = \theta r$ from the jet axis perpendicular to the direction to the
observer.
\label{fig:jetslice} }
\end{figure}

The line profile can be determined using the geometry
of a slice through the jet, shown in Fig.~\ref{fig:jetslice}.
The angle $\theta$ is the flow
direction of a jet fluid element relative to the jet axis
on a chord at distance
$\rho = \theta r = \theta R / \Theta$ from the jet axis
perpendicular to the direction to the observer.
The strength of the emission line at this velocity relative
to line center is
proportional to the length of the chord: $l = 2 (R^2-\rho^2)^{1/2}
= 2 R (1 - (\theta/\Theta)^2)^{1/2}$.
The line profile (given as line-of-sight velocity profiles) is

\begin{equation}
I({\rm v}) = I_0 ( 1 - ({\rm v}/{\rm v}_m)^2 ) ^{1/2} ,
\end{equation}

\noindent
The line profile is well approximated by a Gaussian by matching
the full widths at half maximum: $2.35 \sigma$ = $3^{1/2} v_m$,
giving v$_m$ = 989 $\pm$ 47 km s$^{-1}$ and
$\Theta$ = 0.61\arcdeg\ $\pm$ 0.03\arcdeg.

\section{Modelling the Jet Emission Line Fluxes}

\label{sec:models}

Radiative recombination features are practically
nonexistent.  Two weak features are consistent with radiative
recombination continua (RRC) from Ne {\sc X} and Ne {\sc IX}
(see Table~\ref{tab:linefluxes}) but we do not detect the RRC
of other ions with prominent recombination lines:
Mg {\sc XII}, Si {\sc XIII}, Si {\sc XIV},
S {\sc XV}, and S {\sc XVI}.  At high temperatures, the
RRC could be significantly broadened, making them nearly
indistinguishible from the continuum.  At the temperatures
we derive for the gas based on line ratios and detailed fits
to line fluxes, S and Si RRC should be broadened by $\sim 0.3$\AA,
so they should be detectable if photoionization was
a significant process.  Finally, 
the He-like triplets are dominated by the resonance lines, so we
conclude that photoionization is negligible and that the
plasma is collisionally ionized.

Spectral analysis was performed using {\tt ISIS} using
the APED atomic data base of line emissivities and ionization
balance.  The observed flux of an emission
line from a region with electron density $n_e$ and temperature $T$
is given by

\begin{equation}
\label{eq:emline}
f_i  = \frac{ J_i(T) \int n_e^2 dV} { 4 \pi D^2  }
= \frac{ J_i(T) EM(T) } { 4 \pi D^2 }
\end{equation}

\noindent
where $f_i$ is the observed flux of ion transition $i$ in photons
cm$^{-2}$ s$^{-1}$, $J_i(T)$ is
the emissivity of a thin thermal plasma in transition $i$
in photons cm$^{3}$ s$^{-1}$ and cosmic abundances,
$EM(T)$ is the emission measure of material with
temperature $T$, and all quantities are measured in the rest frame of
the plasma.  For the conical jet geometry shown in
Fig.~\ref{fig:jetgeometry} where the temperature varies
with distance $r$, we approximate equation~\ref{eq:emline} by
a sum over a discrete set of independent sections defined by
$0.5 r_j < r < 1.5 r_j$:

\begin{equation}
\label{eq:emr3}
f_i  =  \frac{ \Omega \sum_j J_i(T_j) n_j^2 r_j^3 } { 4 \pi D^2  }
\end{equation}

\noindent
where $\Omega = \pi \Theta^2$ and $j$ indicates the component
of a multi-temperature model where the average electron
density is $n_j$ and the average electron temperature is $T_j$.

Using the ratios of the H-like to He-like line fluxes,
we can estimate the temperature of the gas that produces
most of each element's emission, ignoring, for now,
the contributions to other emission lines.  Using emissivities
from the APED data base, we find temperatures in the
range from 1.2 to 7.6 keV, shown in table~\ref{tab:lineratios}.
Emission measures, $EM_{1T}$, are also derived, assuming
cosmic abundances and that all
emission from each ion originates in gas of a single temperature
given by the line ratio.  The results show
that there is a broad range of temperatures in the outflow
ranging from 1 to 10 keV ($10^{7-8}$ K) and that each
element's emission lines sample a different region in the jet.
These temperature estimates are independent of the
elemental abundances and are approximately independent
of the line-of-sight column density because the wavelengths
of the H-like and He-like lines are similar so that the
ISM opacities are nearly the same.
The emission measure
can be overestimated due to the contributions from
regions of the jet at other temperatures and requires
an estimate of the line of sight $N_H$.

The blue Si {\sc XIII} and S {\sc XV} forbidden and
intercombination lines are somewhat blended with
the broadened resonance lines but are discernably weaker.
The multi-Gaussian line fits
were used to determine the density- and temperature-sensitive
line ratios defined by \citet{porquet00}: $R(n_e) = f/i$ and
$G(T_e) = (i+f)/r$, where $r$, $i$, and $f$ are the strengths
of the resonance, intercombination, and forbidden lines, respectively.
For Si {\sc XIII}, we find $R = 1.18 \pm 0.26$ and
$G = 0.92 \pm 0.13$ and for S {\sc XV}, we obtain $R = 11 \pm 24$ and
$G = 0.65 \pm 0.20$.  The S {\sc XV} line does
not give a good constraint on the density so we rely on the Si {\sc XIII}
fit results.  Using curves from \citet{porquet00}, the value of
$G$ indicates that $T_e > 3 \times 10^6$ K, which is
consistent with the estimate of $1.9 \times 10^7$ K given in
table~\ref{tab:lineratios}.  At a $T = 10^7$ K, the value of
$R$ gives of $n_e \sim 10^{14}$ cm$^{-3}$.
Using the estimate of the emission measure from table~\ref{tab:lineratios}
and assuming $N_H = 10^{22}$ cm$^{-2}$,
we now have an estimate of the emission volume for gas at
a temperature of $\sim 10^7$ K: $2 \times 10^{30}$ cm$^3$.

For a conical geometry with the opening angle derived from the
line widths, then we obtain $r \sim 2 \times 10^{11}$ cm.  Note
that $r$ is the distance from the {\em apex} of a cone, which
may be truncated at the accretion disk.  The jet flow takes
about 20 s to reach this distance at 0.27$c$, which is much
longer than the recombination time at the estimated density,
$\sim 10^{-3}$ s, so ionization balance can be achieved in the flow.
There is X-ray emission out to 5\arcsec\ from the core at a scale
of $3.6 \times 10^{17}$ cm but we have not yet detected emission
lines at this distance.

\begin{figure}
\epsscale{0.6}
\plotone{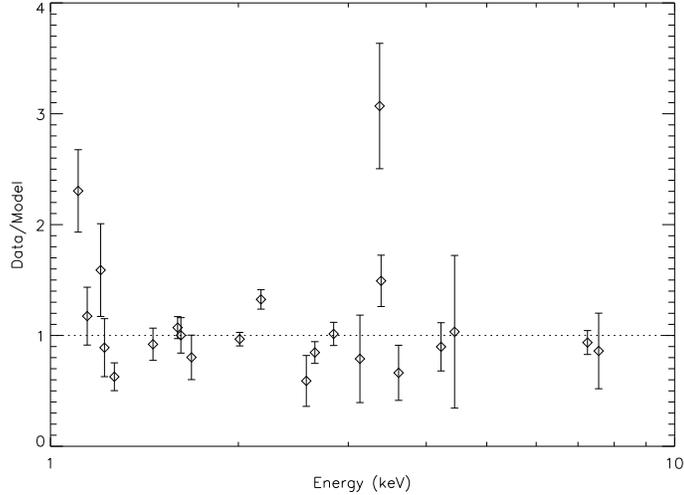}
\caption{Ratio of the observed line flux to that predicted from
the four component thermal model given in table~\ref{tab:emfit}.
Most line fluxes agree with the model to within 50\%.
\label{fig:linefits} }
\end{figure}

A more refined estimate of the emission measure distribution can
be obtained by fitting the line flux data to a multi-temperature
emission model assuming solar abundances.
We obtained a moderately good fit to the line flux
data with a model with four components, as given in table~\ref{tab:emfit}
and shown in Fig.~\ref{fig:emeasure}.
The ISM absorption was fitted simultaneously, giving
$N_H = 2.20 \pm 0.07 \times 10^{22}$ cm$^{-2}$, substantially
larger than estimated from the simplistic power law fit.
The ratios of the
line fluxes to the values predicted from the four component
model are shown in Fig.~\ref{fig:linefits}.
There are indications that the
spectrum is somewhat more complex than our model indicates:
there may be additional lines at 8.65 \AA\ and 9.15 \AA\ that
are not included in the model, the lines at 7.3 \AA\ (Mg {\sc XI}
1s3p-1s$^2$) and 9.8 \AA\ (Fe {\sc XXIV}) appear to be somewhat
stronger in the model than in the data, and the line at
11.2\AA\ (Ne {\sc X}, Fe {\sc XXIII}) is $\times 2$ stronger
than predicted by the model.
Almost all of the other line fluxes are
explained to within 50\% and the strongest lines are modelled
to within 10-20\%.  We consider this to be a satisfactory
agreement, given the poor statistics in the long wavelength
portion of the spectrum which makes it difficult to measure
weak lines from gas in the jet which might be cooler than
$6 \times 10^6$ K.

The continuum expected from the four component model of the blue jet
is comparable to the observed continuum, so the abundances of
Fe, S, Si, Mg, and Ne were
increased by 30\% relative to H and He
in order to allow for continuum from the
red jet and so that the observed continuum is not exceeded in
the 4-7\AA\ region.  The overall fit to the data is very good,
as shown in Figs.~\ref{fig:spectra1} to \ref{fig:spectra3}.
In the 8-12\AA\ region, the model predicts a continuum that
is systematically low by up to a factor of two.  An additional
continuum component is required.
With such a large contribution from thermal bremsstrahlung,
any nonthermal emission that might arise in an accretion disk
must be fainter than $\sim 10^{34}$ erg s$^{-1}$.

Weighted by the emission measures from detailed fits
to emission line fluxes, the Si {\sc XIII} emission is dominated
by gas at a temperature of $\sim 1.3 \times 10^7$ K, which is
close to the temperature of the second component given in
table~\ref{tab:emfit}.
At $1.3 \times 10^7$ K, the observed value of $R$ gives log $n_e$ = 14.0 $\pm$
0.2.  Thus, we assign a density
of $10^{14}$ cm$^{-3}$ to the second emission measure component.  This
component's radius is estimated from Eq.~\ref{eq:emr3} which
gives $r = 1.22 \times 10^{11}$ cm.  Assuming
an asymptotic form of the adiabatic cooling of an freely expanding
jet, $T \propto r^{-4/3}$ (e.g. Eq. 4 from the paper by Kotani et al.),
then we can estimate the radius of each emission component and,
using Eq.~\ref{eq:emr3}, the electron density at that radius.
The results
are given in table~\ref{tab:emfit}.  For a uniform outflow
at constant velocity, $n \propto r^{-2}$, so we expect

\begin{equation}
EM(T) \propto n^2 r^3 \propto r^{-1} \propto T^{3/4} .
\end{equation}

\noindent
This line is plotted in Fig.~\ref{fig:emeasure}, normalized
to provide a good fit for $T < 5 \times 10^7$ K.
The emission measure at the highest
temperature deviates from the expected relation by a factor of 2
but the two remaining measurements agree very well.

The jet mass outflow rate, $\dot{m}$, is computed from

\begin{equation}
\dot{m} = \pi * (r \Theta)^2 {\rm v}_j (1+X) \mu n_e m_p ,
\end{equation}

\noindent
where $X$ is the ratio of the total
ion density to $n_e$ and is about 0.92 for completely ionized
gas at cosmic abundances, $\mu$ is the mean molecular weight
for this gas and $m_p$ is the mass of the proton.
We find $\dot{m} = 1.5 \times 10^{-7} ~ M_{\sun}$ yr$^{-1}$,
derived primarily from the three low temperature components.
With our measurement of the jet velocity, we have a value
for the jet kinetic luminosity: $3.2 \times 10^{38}$ erg s$^{-1}$,
which is $\times$1000 larger than the
unabsorbed 2-10 keV X-ray luminosity.

For the red jet, a simple analysis of the H- and He-like
line flux ratios for Fe and Si gives a two component model
with $EM(T)$ values significantly lower than estimated for
the blue jet, even after correcting for ISM using the higher
$N_H$ estimate and accounting for Doppler flux diminution.
The results, given in table~\ref{tab:lineratios},
are rather uncertain due to the weak Fe {\sc XXVI}
line and the blended Si {\sc XIII} lines.  Tentatively, it
appears that the emission measures are 20-25\% of those
found for the blue jet.  \citet{kotani} obtained a similar
result from {\em ASCA} data, obtaining a fraction of about 35\%
for just the Fe {\sc XXV} line ratio.  \citet{kotani} cited
two effects that would reduce the Fe line flux from the red
jet relative to the blue jet: occultation of the hot
inner zones by an accretion
disk and attenuation by additional cold gas along the line
of sight to the more distant red jet.  Although the
statistics are poor, the Fe {\sc XXVI} to Fe {\sc XXV} ratios are
consistent: 0.30 $\pm$ 0.17 for the blue jet and
0.17 $\pm$ 0.11 for the red jet, so we find that
the red and blue jet temperatures are similar.  A longer
observation is needed to obtain a better detection of the
red Fe {\sc XXVI} line.
As to the attenuation hypothesis, we obtain
about the same ratio of emission
measures for lines near 2\AA\ as those near 7.5\AA, which
would be difficult to explain in a scenario involving absorption by
neutral gas.

\section{Discussion and Summary}

The X-ray lines are
emitted in an optically thin region where $T$ drops
from $1.1 \times 10^8$ K to $6 \times 10^6$K, $n_e$ drops
from $2 \times 10^{15}$
to $4 \times 10^{13}$ cm$^{-3}$, at a distances
up to $2 \times 10^{11}$ cm from the jet base.
The X-ray continuum is dominated
by the thermal bremsstrahlung from the jet so any nonthermal emission
from the core must be $< 10^{34}$ erg s$^{-1}$ unless the abundances
in the jet are substantially non-solar.
Weak extended X-ray emission
is observed out to $\sim 10^{17}$ cm but the gas is likely to be too
cool to emit X-rays thermally.

Line profile measurements are consistent with the interpretation
that the emission lines originate in conical jets with constant
opening angle and constant flow speed.
The lines are not double peaked, which would be expected if the
gas were confined to the rim of the conical outflow, as might
occur if the heavy elements were entrained from 
gas surrounding a leptonic jet.  The lines
are not skewed, as might be expected if material was accelerated
through a narrow ``nozzle''.
From the consistency and narrowness
of the line widths, we have a very accurate measurement of
the jet opening angle: 1.23 $\pm$ 0.06\arcdeg.
The optical lines give a
somewhat larger value which may indicate that the optical
emission results from interactions of the rapidly moving jet
with ambient material much further along the jet,
as suggested by \cite{begelman}.
The jet velocity we measure is 0.2699 $\pm 0.0007 c$, which
is larger than the velocity inferred from optical emission lines
by 2920 $\pm$ 440 km s$^{-1}$, lending support to the
hypothesis that the optical and X-ray line emission regions
are physically distinct.

The blueshifted radiative recombination continua of Ne {\sc X}
and Ne {\sc IX} indicate that there is a region of cooler
photoionized plasma in the jet.  The line widths give a temperature
estimate of about $< 10^5$ K, which would be found in the
jet at $> 4 \times
10^{13}$ cm from the compact object, assuming the adiabatic
expansion model.  Correcting for interstellar absorption, the
ratio of the Ne {\sc IX} to Ne {\sc X} fluxes is 2.1 $\pm$ 0.6,
giving an ionization parameter of $\xi \equiv L_x/(n_e r^2) \approx 40$.
For the constant velocity jet, $n_e r^2$ is constant at
about 1.7 $\times 10^{36}$ cm$^{-1}$
which would require a photoionizing power of $7 \times 10^{39}$
erg s$^{-1}$, greatly exceeding both the observed X-ray luminosity and
the kinetic power in the jet.
Even if this luminosity were present but somehow unobserved,
it would still be difficult to understand why there are no
detectable Mg or Si RRC.

As pointed out by \citet{kotani}, the electron scattering
optical depth across the base of the jet should not be significantly
larger than one because otherwise the emission lines would be
significantly broadened by Comptonization,  which would broaden
lines by $\delta \lambda \simeq \lambda (kT/m_e c^2)^{1/2)} = $
0.26 \AA\ for the Si resonance lines at about 6 \AA.
This level of broadening is not observed in the S or Si resonance
lines where it
would dominate the velocity broadening or produce a broad pedestal.
For the innermost component of the jet emission model, we can
estimate the electron scattering optical
depth, $\tau_{es} = n_e \sigma_T \Theta r = 0.3$, or marginally optically
thin.
Since $n_e \propto r^{-2}$,
$\tau \propto r^{-1}$ so the rest of the jet must be optically
thin to electron scattering as well.

\begin{figure}
\epsscale{0.6}
\plotone{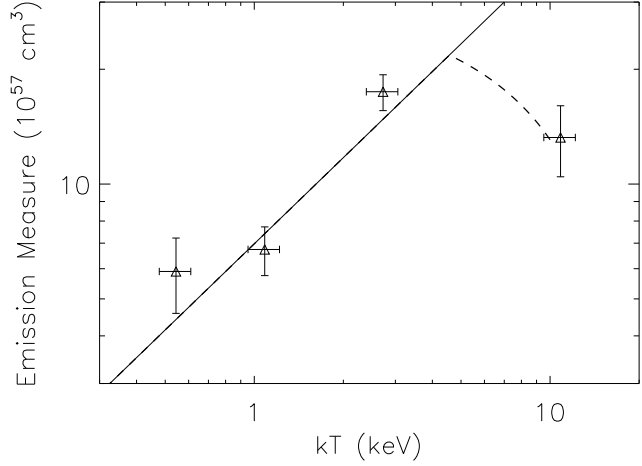}
\caption{Emission measure $vs.$ temperature from the
multi-temperature fits to the blue jet line fluxes.
The vertical error bars are determined from the
fit while the horizontal error bars denote the coarseness of the
temperature grid used. The line gives the relation expected
for an adiabatically expanding uniform outflow in the asymptotic
limit.  The high $T$ point is low by a factor of 2, which may
result from increased optical thickness or perhaps truncation
of the jet.
The dashed line gives a model of the emission measure that accounts
for the decreased volume of a jet that is truncated at
$r = 2.1 \times 10^{10}$
cm with a maximum temperature of $1.1 \times 10^8$ K.
\label{fig:emeasure} }
\end{figure}

The optical depth in the resonance lines is
be significantly larger, however, and might explain the drop in emission
measure at the highest temperatures seen in figure.  In the region that emits the Si {\sc XIII}
resonance line, for example, we find that the resonant optical depth
at line center is $\tau_0 \approx 8$, while $\tau_{es} \approx 0.08$.
This means that resonance scattering would increase the effective path
length of a line photon, thereby increasing the probability that it is Compton
scattered out of the line core.  
The effective optical depth to
line escape is approximately  $\tau_{eff} \approx (\tau_0 \tau_{es})^{1/2}$,
(see \citet{fr72} and \citet{fra72}).
For Si {\sc XIII}, $\tau_{eff} \sim 0.8$, so Compton
scattering may be marginally important (although a proper radiative
transfer calculation using the jet geometry and velocity structure 
is warranted).
In the hottest region of table~\ref{tab:emfit} $\tau_{eff} \approx 1.5$ 
for the strongest resonance lines of Fe {\sc XXV} and Fe {\sc XXVI}, both
of which have $\tau_0 \approx 8$. Using
instead the density for this temperature expected by the adiabatic relation 
(Fig.~\ref{fig:emeasure}) would give $\tau_{eff} \approx 2.5$ for these lines.
Therefore, one
explanation for why the emission measure appears to deviate
from the expected adiabatic relation (Fig.~\ref{fig:emeasure})
is that the emission lines from the highest density regions
are Compton scattered from the core due to resonance trapping, 
and simply contribute to the continuum. 

A reduced emission measure at high temperature might also be
expected if the cone of the jet is truncated
at $r_0 = 2.0 \times 10^{10}$ cm, so that the
emission measure is integrated over a smaller volume than assumed.
In this truncation model, $n_e$ increases by $\times$ 2
for the highest temperature component in table~\ref{tab:emfit} to
about $4 \times 10^{15}$ cm$^{-3}$.  A revised emission measure
model is also shown in Fig.~\ref{fig:emeasure} and agrees
with the data.
The radius of the base of the visible jet is
$\Theta r_0 = 2.3 \times 10^8$ cm.

We note the interesting agreement between the expansion
velocity of the jet perpendicular to the jet axis,
v$_j \sin \Theta$ = 866 $\pm$ 41 km s$^{-1}$, and
the sound speed in the rest frame of the flowing gas,

\begin{equation}
c_s = [ \frac{5 k T}{3 \mu m_p (1+X)} ]^{1/2}   ,
\end{equation}

\noindent
which is 1149 km s$^{-1}$ for our estimate of the jet
base temperature, $T = 1.1 \times 10^8$ K.  We speculate
that this agreement is physical: the jet expands
at the sound speed of hydrogen and that the heavier elements
are coupled to the protons at the base of the jet.
The coupling process is unknown but could be related
to the jet acceleration process as both must affect
protons and ions of high Z elements equally.

\acknowledgments

We thank the referee for many insightful comments on the
submitted manuscript, especially for pointing out the
importance of resonant line opacity.
This work has been supported in part under NASA contracts NAS8-38249
and SAO SV1-61010.

\clearpage

\clearpage

\begin{deluxetable}{clccl}
\tablecolumns{5}
\tablewidth{0pc}
\tabletypesize{\scriptsize}
\tablecaption{SS 433 X-ray Emission Lines \label{tab:linefluxes} }
\tablehead{\colhead{$\lambda_{rest}$} & \colhead{$\lambda_{obs}$}
    & \colhead{$z$} & \colhead{Flux} & \colhead{Identification} \\
  \colhead{(\AA)} & \colhead{(\AA)} & \colhead{}
    & \colhead{($10^{-6}$ ph cm$^{-2}$ s$^{-1}$)} & \colhead{} }
\startdata
\cutinhead{Blue Jet}
1.780 & 1.641 $\pm$ 0.002 & -0.0779 $\pm$ 0.0012 & 111 $\pm$ 44 & Fe {\sc
XXVI} Ly$\alpha$ \tablenotemark{a} \\
1.855 & 1.710 $\pm$ 0.001 & -0.0781 $\pm$ 0.0005 & 373 $\pm$ 43 & Fe {\sc XXV} 1s2p-1s$^2$\\
3.020 & 2.795 $\pm$ 0.006 & -0.0745 $\pm$ 0.0021 & 12 $\pm$ 8 & Ca {\sc XX} Ly$\alpha$\\
3.186 & 2.935 $\pm$ 0.004 & -0.0788 $\pm$ 0.0013 & 34 $\pm$ 8 & Ca {\sc XIX} 1s2p-1s$^2$\\
3.733 & 3.432 $\pm$ 0.006 & -0.0805 $\pm$ 0.0015 & 16 $\pm$ 6 & Ar {\sc XVIII} Ly$\alpha$\\
3.952 & 3.650 $\pm$ 0.003 & -0.0764 $\pm$ 0.0009 & 20 $\pm$ 6 & Ar {\sc XVII} 1s2p-1s$^2$\\
3.991 & 3.682 $\pm$ 0.002 & -0.0774 $\pm$ 0.0005 & 38 $\pm$ 7 &  S {\sc XVI} Ly$\beta$\\
4.299 & 3.959 $\pm$ 0.006 & -0.0791 $\pm$ 0.0013 & 12 $\pm$ 6 & S {\sc XV} 1s3p-1s$^2$\\
4.729 & 4.362 $\pm$ 0.002 & -0.0776 $\pm$ 0.0003 & 97 $\pm$ 10 & S {\sc XVI} Ly$\alpha$\\
5.055 & 4.675 $\pm$ 0.004 & -0.0778 $\pm$ 0.0003 & 121 $\pm$ 14 & S {\sc
XV} 1s2p-1s$^2$ \tablenotemark{d} \\
5.217 & 4.825 $\pm$ 0.005 & -0.0752 $\pm$ 0.0010 & 18 $\pm$ 7 & Si {\sc XIV} Ly$\beta$\\
6.182 & 5.703 $\pm$ 0.002 & -0.0776 $\pm$ 0.0002 & 182 $\pm$ 12 & Si {\sc XIV} Ly$\alpha$\\
6.675 & 6.169 $\pm$ 0.003 & -0.0779 $\pm$ 0.0002 & 158 $\pm$ 10 & Si {\sc
XIII} 1s2p-1s$^2$ \tablenotemark{d} \\
7.986 & 7.370 $\pm$ 0.005 & -0.0771 $\pm$ 0.0006 & 16 $\pm$ 4 & Fe {\sc XXIV} \\
8.310 & 7.663 $\pm$ 0.004 & -0.0779 $\pm$ 0.0005 & 25 $\pm$ 4 & Fe {\sc XXIII/XXIV} \\
8.421 & 7.756 $\pm$ 0.002 & -0.0789 $\pm$ 0.0003 & 54 $\pm$ 5 & Mg {\sc
XII} Ly$\alpha$ \tablenotemark{b} \\
9.102 & 8.370 $\pm$ 0.004 & -0.0803 $\pm$ 0.0004 & 26 $\pm$ 5 & Ne {\sc X} RRC
\tablenotemark{c} \\
9.181 & 8.495 $\pm$ 0.018 & -0.0747 $\pm$ 0.0020 & 38 $\pm$ 6 & Mg {\sc XI} 1s2p-1s$^2$\\
10.368 & 9.552 $\pm$ 0.005 & -0.0788 $\pm$ 0.0004 & 24 $\pm$ 4 & Ne {\sc IX} RRC
\tablenotemark{c} \\
10.634 & 9.795 $\pm$ 0.006 & -0.0789 $\pm$ 0.0005 & 20 $\pm$ 4 & Fe {\sc
   XXIV}\tablenotemark{e} \\
11.008 & 10.159 $\pm$ 0.010 & -0.0771 $\pm$ 0.0009 & 17 $\pm$ 5 & Fe {\sc
   XXIV}\tablenotemark{e} \\
11.176 & 10.303 $\pm$ 0.007 & -0.0781 $\pm$ 0.0006 & 19 $\pm$ 5 & Fe {\sc XXIV} \\
11.736 & 10.827 $\pm$ 0.007 & -0.0775 $\pm$ 0.0006 & 18 $\pm$ 4 & Fe {\sc XXIII} \\
12.134 & 11.194 $\pm$ 0.004 & -0.0783 $\pm$ 0.0003 & 31 $\pm$ 5 & 
	Ne {\sc X} Ly$\alpha$ \& Fe {\sc XXIII}\\
\cutinhead{Rest frame}
1.937 & 1.939 $\pm$ 0.002 & 0.0010 $\pm$ 0.0008 & 49 $\pm$ 14 & Fe {\sc I} \\
\cutinhead{Red Jet}
1.592 & 1.836 $\pm$ 0.003 & 0.1538 $\pm$ 0.0016 & 55 $\pm$ 16 & Ni {\sc XXVII}  1s2p-1s$^2$ \\
1.780 &  2.057 $\pm$ 0.005 & 0.1560 $\pm$ 0.0026 & 22 $\pm$ 10 & Fe {\sc XXVI} Ly$\alpha$\\
1.855 &  2.147 $\pm$ 0.001 & 0.1577 $\pm$ 0.0006 & 129 $\pm$ 16 & Fe {\sc XXV}  1s2p-1s$^2$ \\
6.182 & 7.133 $\pm$ 0.004 & 0.1537 $\pm$ 0.0006 & 34 $\pm$ 4 & Si {\sc XIV} Ly$\alpha$\\
6.675 & 7.747 $\pm$ 0.018 & 0.1610 $\pm$ 0.0026 & 25 $\pm$ 4 & Si {\sc
XIII} 1s2p-1s$^2$ \tablenotemark{b} \\
7.986 & 9.214 $\pm$ 0.006 & 0.1538 $\pm$ 0.0007 & 27 $\pm$ 4 & Fe {\sc XXIV} \\
\enddata

\tablenotetext{a}{Only HEG data were used.}
\tablenotetext{b}{Mg {\sc XII} from the blue jet is somewhat confused with
Si {\sc XIII} from the red jet.  There is also an Fe {\sc XXIV}
line from the blue jet near the location of the Si {\sc XIII} resonance line.}
\tablenotetext{c}{Radiative recombination continua.}
\tablenotetext{d}{He-like triplet system fitted with three lines
with independent fluxes and fixed rest wavelengths.}
\tablenotetext{e}{Blend; rest wavelength is based on a flux-weighted average.}
\end{deluxetable}

\begin{deluxetable}{ll}
\tablecolumns{2}
\tablewidth{0pc}
\tablecaption{Line Widths
\label{tab:linewidths} }
\tablehead{\colhead{Wavelength Range} & \colhead{$v$\tablenotemark{a}} \\
   \colhead{(\AA)} & {(km s$^{-1}$)} }
\startdata
1.5 -- 2.5	&	892 $\pm$  162 \\
2.5 -- 4.0	&	741 $\pm$  138 \\
4.0 -- 5.5	&	742 $\pm$  ~75 \\
5.5 -- 7.5	&	810 $\pm$  ~61 \\
7.5 -- 10.0 &	656 $\pm$  ~75 \\
10.0 -- 11.5 &	595 $\pm$  ~86 \\
\enddata
\tablenotetext{a}{Velocity width ($\sigma$) of Gaussian line profiles.}
\end{deluxetable}

\begin{deluxetable}{lcccc}
\tablecolumns{5}
\tablewidth{0pc}
\tablecaption{Estimated Temperatures and Emission Measures
	from Line Ratios using Single Temperature Models
\label{tab:lineratios} }
\tablehead{
  \colhead{} & \multicolumn{2}{c}{Blue Jet} & 
	\multicolumn{2}{c}{Red Jet} \\
  \cline{2-3} \cline{4-5} \\
  \colhead{Element} & \colhead{$T$} & \colhead{$EM_{1T}$\tablenotemark{a}}
    & \colhead{$T$} & \colhead{$EM_{1T}$\tablenotemark{a}} \\
  &  (10$^6$ K) & (10$^{57}$ cm$^3$) & (10$^6$ K) & (10$^{57}$ cm$^3$) }
\startdata
Fe	&	88  & 20.1  & 71 & 3.9 \\
Ca	&	41  & 26.1  & \nodata & \nodata \\
Ar	&	23  & 27.0  & \nodata & \nodata \\
S	&	26  & 21.0  & \nodata & \nodata \\
Si	&	19 & 18.5  & 20 & 2.4 \\
Mg	&	13 &  8.0  & \nodata & \nodata \\
\enddata

\tablenotetext{a}{Emission measure for a single temperature model
that reproduces that elements' H- and He-like emission line 
fluxes at a distance of 4.85 kpc, for an interstellar absorption
column density of $10^{22}$ cm$^{-2}$, and corrected for
Doppler shifts.}
\end{deluxetable}

\begin{deluxetable}{cccc}
\tablecolumns{4}
\tablewidth{0pc}
\tablecaption{Blue Jet Parameters from a Multi-Temperature Model\tablenotemark{a}
\label{tab:emfit} }
\tablehead{\colhead{$T$} & \colhead{$EM$} & \colhead{$r$} & \colhead{$n_e$} \\
  (10$^6$ K) & (10$^{57}$ cm$^3$) & (10$^{10}$ cm) & (10$^{14}$ cm$^{-3}$) }
\startdata
   6.3 &  5.90 $\pm$  1.32 & 20.5  &  0.45 \\
  12.6 &  6.57 $\pm$  1.00 & 12.2  &  1.00 \\
  31.6 & 18.   $\pm$  1.92 &  6.13 &  4.66 \\
 126.  & 12.8  $\pm$  2.87 &  2.17 & 18.6  \\
\enddata
\tablenotetext{a}{The interstellar absorption
column density was fit simultaneously with the $EM$ values
at each temperature.  The best fit absorption column was
$2.20 \pm 0.07 \times 10^{22}$ cm$^{-2}$.  The temperatures were set
to values on a logarithmic grid which minimized the $\chi^2$.}
\end{deluxetable}


\begin{thebibliography}{}
\bibitem[Abell \& Margon(1979)]{am79}
	Abell, G.\ O. \& Margon, B.\ 1979, Nature, 279, 701
\bibitem[Begelman et al.(1980)]{begelman}
	Begelman, M.\ C., Sarazin, C.\ L., Hatchett, S.\ P.,
	McKee, C.\ F., and Arons, J.\ 1980, \apj, 238, 722
\bibitem[Brinkmann et al.(1991)]{brinkmann91}
	Brinkmann, W., Kawai, N., Matsuoka, M., and
	Fink, H.\ H.\ 1991, \aap, 241, 112
\bibitem[Canizares et al.(2001)]{canizares00}
	Canizares et al.\ 2001, in preparation
\bibitem[Davidson \& McCray(1980)]{dm80}
	Davidson \& McCray, 1980, \apj, 240, 1082
\bibitem[Fabian \& Rees(1979)]{fr79}
	Fabian, A.\ C.\ \& Rees, M.\ J.\ 1979, \mnras, 187, 13P
\bibitem[Felton \& Rees(1972)]{fr72}
	Felton, J.\ E.\ \& Rees, M.\ J.\ 1972, \aap, 17, 226
\bibitem[Felton, Rees \& Adams (1972)]{fra72}
	Felton, J.\ E.\, Rees, M.\ J.\ \& Adams, T.\ F.\ 1972, \aap, 21, 139
\bibitem[Gladyshev et al.(1987)]{gladyshev}
	Gladyshev, S.\ A., Goranskii, V.\ P., and
	Cherepashchuk, A.\ M.\ 1987, Sov. Astron., 31, 541
\bibitem[Hjellming \& Johnston(1981)]{hjellming}
	Hjellming, R. \& Johnston, S.\ 1981, \apjl, 246, L141
\bibitem[Kotani et al.(1996)]{kotani}
	Kotani, T., Kawai, N., Matsuoka, M., and
	Brinkmann, W.\ 1996, \pasj, 48, 619
\bibitem[Margon et al.(1977)]{margon77}
	Margon et al.\ 1977, \apjl, 246, L141
\bibitem[Milgrom(1979)]{milgrom79}
	Milgrom, M.\ 1979, \aap, 76, L3
\bibitem[Margon (1984)]{margon84}
	Margon, B.\ 1984, \araa, 22, 507
\bibitem[Margon \& Anderson(1989)]{ma89}
	Margon, B., \& Anderson, S.\ F.\ 1989, \apj, 347, 448
\bibitem[Marshall et al.(1979)]{marshall79}
	Marshall, F.\ E., Swank, J.\ H., Boldt, E.\ A., Holt, S.\ S.
	Serlemitsos, P.\ J.\ 1979, \apj, 230, 145
\bibitem[Porquet \& Dubau(2000)]{porquet00}
	Porquet, D., \& Dubau, J.\ 2000, \aap, 143, 495
\bibitem[Vermeulen et al.(1993)]{vermeulen}
	Vermeulen, R.\ C., Schilizzi, R.\ T., Spencer, R.\ E.,
	Romney, J.\ D., and Fejes, I.\ 1993, \aap, 270, 177
\bibitem[Watson et al.(1986)]{watson}
	Watson, M.\ G., Stewart, G.\ C., Brinkmann, W.,
	and King, A.\ R.\ 1986, \mnras, 222, 261
\bibitem[Watson et al.(1983)]{watson83}
	Watson, M.\ G., Willingale, R.,
	Grindlay, J.\ E., and Seward, F.\ D.\ 1983, \apj, 273, 688
\end{thebibliography}
\end{document}